\documentclass[11pt]{article}

\usepackage[draft]{graphics}
\usepackage{graphicx}          	 % for eps figures
\usepackage{bm}                 % for bold math symbols
\usepackage{amsmath}             % for \text and such\gstar
\usepackage{amssymb}
\usepackage{amsfonts}             
\usepackage{verbatim}           % useful for commenting out stuff
\usepackage{amsthm}             % does theorems etc. nicely, but

\usepackage{mathtools}
\usepackage{relsize}

\usepackage[colorlinks,citecolor=blue]{hyperref}

\usepackage{makeidx}

\theoremstyle{plain}             % This is the default
\newtheorem{theorem}{Theorem}[section]

\newtheorem{proposition}[theorem]{Proposition}
\theoremstyle{definition}
\newtheorem{example}[theorem]{Example}

\newtheorem{remark}[theorem]{Remark}

\makeatletter
\@addtoreset{equation}{chapter}

\makeatother

\def\eqref#1{(\ref{#1})}
\def\dsp{\displaystyle}
\def\Frac#1#2{\frac
{
 {\raise.6ex
 \hbox{$\displaystyle#1$}}
}
{
 {\lower.6ex
 \hbox{$\displaystyle#2$}}
 }
}

\numberwithin{equation}{section}

\def\i{{i}}

\def\bigOxe{\sqcup \kern-2.3mm \sqcap}

\def\eoexample{{\unskip\nobreak\hfil\penalty50	
\hskip2em\hbox{}\nobreak\hfil$\diamondsuit$
\parfillskip=0pt\finalhyphendemerits=0\medbreak}}%TeXbook p. 106 (signed)

\def\eoremark{{\unskip\nobreak\hfil\penalty50	
\hskip2em\hbox{}\nobreak\hfil$\triangle$
\parfillskip=0pt\finalhyphendemerits=0\medbreak}}%TeXbook p. 106 (signed)

\makeatother

\def\dsp{\displaystyle}
\def\Frac#1#2{\frac
{
 {\raise.6ex
 \hbox{$\displaystyle#1$}}
}
{
 {\lower.6ex
 \hbox{$\displaystyle#2$}}
 }
}

\def\sign{{\rm sign}}

\def\CHFs#1#2#3{
\ {}_1F_1\left({a};{c};{z}\right)
}

\def\binom#1#2{
\renewcommand{\arraystretch}{0.9}
\left(
\begin{array}{c}
\begin{array}{c}\hskip-10pt#1\end{array}\\
\begin{array}{c}\hskip-10pt#2\end{array}
\end{array}
\hskip-10pt
\renewcommand{\arraystretch}{1.0}
\right)}

\def\bigO{{\cal O}}
\def\calC{{{\cal C}}}

\def\tfrac#1#2{{{\lower.6ex
\hbox{$\scriptstyle#1$}}\over 
{\raise.7ex
\hbox{$\scriptstyle#2$}}}}

\def\sign{{\rm sign}}

\def\tfrac#1#2{{{\lower.6ex
\hbox{$\scriptstyle#1$}}\over 
{\raise.7ex
\hbox{$\scriptstyle#2$}}}}

\def\eoexample{{\unskip\nobreak\hfil\penalty50	
\hskip2em\hbox{}\nobreak\hfil$\diamondsuit$
\parfillskip=0pt\finalhyphendemerits=0\medbreak}}%TeXbook p. 106 (signed)

\def\eoremark{{\unskip\nobreak\hfil\penalty50	
\hskip2em\hbox{}\nobreak\hfil$\triangle$
\parfillskip=0pt\finalhyphendemerits=0\medbreak}}%TeXbook p. 106 (signed)

\begin{document}

% \title{Entropic uncertainty measures for $D$-dimensional \\ 
% hydrogenic systems at large $D$}
\title{Entropic functionals of Laguerre and Gegenbauer polynomials with large parameters}

\author{
N. M. Temme\\ 
IAA, 1825 BD 25, Alkmaar, The Netherlands\footnote{Former address: CWI, Science Park 123, 1098 XG Amsterdam, The Netherlands}
\\
{nico.temme@cwi.nl}\\
\and
I. V. Toranzo\\
Departamento de F{\i}sica At\'{o}mica, Molecular y Nuclear, \\
Universidad de Granada, Granada 18071, Spain\\ 
{leargrost@gmail.com}\\
\and
J. S. Dehesa\\
 Instituto Carlos I de F{\i}sica Te\'orica y Computacional, \\
 Universidad de Granada, Granada 18071, Spain\\
{dehesa@ugr.es}
}

%\address[label1]{Departamento de F\'{\i}sica At\'{o}mica, Molecular y Nuclear, Universidad de Granada, Granada 18071, Spain\\ Instituto Carlos I de F\'{\i}sica Te\'orica y Computacional, Universidad de Granada, Granada 18071, Spain}
\maketitle
\begin{abstract}
The determination of the physical entropies (R\'enyi, Shannon, Tsallis) of high-dimensional quantum systems subject to a central potential requires the knowledge of the asymptotics of some power and logarithmic integral functionals of the hypergeometric orthogonal polynomials which control the wavefunctions of the stationary states. For the $D$-dimensional hydrogenic and oscillator-like systems, the wavefunctions of the corresponding bound states are controlled by the Laguerre ($\mathcal{L}_{m}^{(\alpha)}(x)$) and Gegenbauer ($\mathcal{C}^{(\alpha)}_{m}(x)$) polynomials in both position and momentum spaces, where the parameter $\alpha$ linearly depends on $D$. In this work we study the asymptotic behavior as $\alpha \to \infty$ of the associated entropy-like integral functionals of these two families of hypergeometric polynomials.

\end{abstract}

%\pacs{89.70.Cf, 89.70.-a, 32.80.Ee, 31.15.-p}


\section{Introduction}
Let us define the integral functionals
\begin{equation}
\label{eq:intro01}
I_{1} (m,\alpha)= \int\limits_{0}^{\infty}x^{\mu-1}e^{-\lambda x} \left|\mathcal{L}_{m}^{(\alpha)}(x)\right|^{\kappa}\,dx, 
\end{equation}
\begin{equation}
\label{eq:intro02}
I_{2}(m,\alpha)= \int_{0}^{\infty} x^{\mu-1}e^{-\lambda x}\bigl(\mathcal{L}^{(\alpha)}_{m}(x)\bigr)^{2}\log\bigl(\mathcal{L}^{(\alpha)}_{m}(x)\bigr)^{2}\, dx,  
\end{equation}
\begin{equation}
\label{eq:intro03}
I_{3}(m,\alpha)  = \int_{-1}^{1} (1-x)^{c\alpha+a} (1+x)^{d\alpha+b}\left|\mathcal{C}^{(\alpha)}_{m}(x)\right|^{\kappa}\,dx, 
\end{equation}
\begin{equation}
\label{eq:intro04}
I_{4}(m,\alpha)= \int_{-1}^{1} (1-x^{2})^{\alpha-\frac{1}{2}}[\mathcal{C}^{(\alpha)}_{m}(x)]^{2}\log \bigl(\mathcal{C}^{(\alpha)}_{m}(x)\bigr)^{2} \, dx,
\end{equation}
where $\mathcal{L}^{(\alpha)}_{m}(x)$ and  $\mathcal{C}^{(\alpha)}_{m}(x)$ are the standard Laguerre and Gegenbauer polynomials, $m=0,1,2,\ldots$, $ \mu>0$, $\lambda >0$,  $\kappa >0$, $c>0$, $d>0$. In certain information-theoretic contexts these integrals are called R\'{e}nyi ($I_{1}, I_{3}$) and Shannon ($I_{2}, I_{4}$) entropic functionals of Laguerre and Gegenbauer polynomials, respectively. This is because they describe the R\'{e}nyi and Shannon information entropies of the probability densities (squared wave functions) which characterize the physical states of the $D$-dimensional quantum systems subject to spherically symmetric potentials. It happens that the solutions (wave functions) of the Schr\"{o}dinger equation of some of these systems are controlled by the Laguerre and Gegenbauer polynomials in the conjugated position and momentum spaces, respectively, being the parameter $\alpha$ of the polynomials a linear function of the dimension $D$.

Indeed, e.g. the wave functions of the stationary states of a three-dimensional single-particle system subject to a central potential $V(\vec{r},t) = V(r)$ are known to have the form $\Psi(\vec{r},t) = \psi(\vec{r})\, e^{-\frac{i}{\hbar}Et}$, where $E$ denotes the state's energy and the corresponding eigenfunction can be expressed in spherical coordinates as 
\begin{equation}
 \psi_{nlm} (\vec{r})=R_{nl}(r)\, Y_{lm} (\theta,\phi)
\end{equation}
with the quantum numbers $n=0,1,2,\ldots$, $l=0,1,2,\ldots$, and
$m=-l,-l+1\ldots,l$. The angular part is given by the spherical harmonics
\begin{equation}
Y_{lm}(\theta,\phi)= \frac{1}{\sqrt{2 \pi}} C_{l-m}^{(l+m)} \left( \cos \theta
\right) \left( \sin \theta \right)^m e^{im \phi}
\end{equation}
with $0\le\theta\le\pi$ and $0\le\phi\le 2\pi$. The symbol $C_k^{(\alpha)}(x)$ denotes the ultraspherical or Gegenbauer polynomials \cite{olver}. The radial part $R_{nl}(r)$ can be often expressed as $\omega^{1/2}(r) y_n(r)$, where $\{y_n(r)\}$ denotes a system of hypergeometric polynomials orthogonal with respect to the weight function $\omega(r)$ in an interval support of the real line. Then, the quantum probability density of the system is given by  
\begin{equation} \label{eq:probdens}
 \rho_{nlm}(\vec{r})  = |\psi_{nlm} \left( \vec{r} \right) |^2 = \omega(r) \left[ y_n(r) \right]^2\,\, \left[ C_{l-m}^{(l+m)}
\left( \cos \theta \right) \right]^2 \left[ \sin \theta \right]^{2m},
\end{equation}
where the radial part is the Rakhmanov density of the polynomials $y_n(r)$, $\omega(r) \left[ y_n(r) \right]^2$, and the angular part is controlled by the Rakhmanov density of the Gegenbauer polynomials $C_{k}^{(\alpha)}(x)$, $\omega^{*}_{\alpha}(x) \left[ C_{k}^{(\alpha)}(x) \right]^2$, where $\omega^{*}_{\alpha}(x)=(1-x^2)^{\alpha-\frac{1}{2}}$ on the interval $\left[-1,+1 \right] $ denotes the associated weight function. In the case of hydrogenic and oscillator-like systems the radial part is given by the Rakhmanov density of the Laguerre polynomials ${\cal{L}}^{(\alpha)}_{m}(r)$ \cite{olver}, $\omega_{\alpha}(r) \left[{\cal{L}}^{(\alpha)}_{m}(r) \right]^2$, where $\omega_\alpha(r)=r^{\alpha} e^{-r}$ on the interval $\left[0,\infty \right)$ is the corresponding associated weight function.

The multiple facets of the spreading of the quantum probability density $\rho_{nlm} (\vec{r})$, which include the intrinsic randomness (uncertainty) and the geometrical profile of the quantum system, can be quantified by means of the dispersion measures (e.g., the variance) and the entropy-like measures  (e.g., R\'enyi, Shannon, Tsallis) of the radial and angular densities.  The variance $V[\rho_{nlm}] = \langle \vec{r}^{2}\rangle - \langle \vec{r} \rangle^{2}  =\langle r^{2}\rangle$ (since $\langle \vec{r} \rangle= 0$ for any central potential) is given by  
\begin{equation} \label{eq:12}
V\left[\rho_{nlm} \right]=\int_{0}^{\infty} r^4 |R_{nl}(r)|^2 dr
%- \left| \int_{0}^{\infty}  r^3 \omega_l(r) \left[y_n(r) \right]^2dr \right|^2,
\end{equation}
and the Shannon entropy  $S[\rho_{n,l,m}] := - \int \rho_{n,l,m}(\vec{r}) \ln \rho(\vec{r}) d\vec{r}$ can be decomposed as $S[\rho_{n,l,m}] = S[R_{n,l}] + S[Y_{l,m}]$, where the radial and angular Shannon entropies are given by
\begin{equation}
 S[R_{nl}]=-\int_{0}^{\infty} |R_{nl}(r)|^2 \ln |R_{nl}(r)|^2 r^2 dr
\end{equation}
and
\begin{equation}
 S[Y_{lm}]=-\int_{0}^{\pi} \sin \theta \,d \theta \int_{0}^{2 \pi} d\phi\,
|Y_{lm}(\theta,\phi)|^2 \ln |Y_{lm}(\theta,\phi)|^2,
\end{equation}
respectively. Moreover the R\'enyi entropies of the quantum state $(n,l,m)$, $\mathcal{R}_{p}[\rho] =  \frac{1}{1-p}\ln \int_{\mathbb{R}^{3}} [\rho(\vec{r})]^{p}\, d\vec{r}, \quad 0<p<\infty,\, p \neq 1,$ can be expressed as
\begin{equation}
\label{eq:renyihyd1}
\mathcal{R}_{p}[\rho_{n,l,m}] = \mathcal{R}_{p}[R_{n,l}]+\mathcal{R}_{p}[Y_{l,m}],
\end{equation}
where $\mathcal{R}_{p}[R_{n,l}]$ denotes the radial part
\begin{equation}
\label{eq:renyi2}
\mathcal{R}_{p}[R_{n,l}] = \frac{1}{1-p}\ln \int_{0}^{\infty} [R_{n,l}(r)]^{2p} r^{2}\, dr,
\end{equation}
and  $\mathcal{R}_{p}[Y_{l,m}]$ denotes the angular part
\begin{equation}
\label{eq:renyi3}
\mathcal{R}_{p}[Y_{l,m}] = \frac{1}{1-p}\ln \int_{0}^{\pi}\sin \theta \,d \theta \int_{0}^{2 \pi} d\phi \,|Y_{l,m}(\theta,\phi)|^{2p}\, .
\end{equation}
Now it is straightforward to see that for three-dimensional hydrogenic and oscillator-like systems the integrals required for the determination of the variance and the angular R\'enyi entropy are of the type $I_1$ given by Eq. (\ref{eq:intro01}), and the integrals involved in the determination of the angular R\'enyi entropy are of the type $I_3$ given by Eq. (\ref{eq:intro03}). Moreover the integrals needed to calculate the radial and angular Shannon entropies belong to the family of functionals $I_2$ and $I_4$ given by Eqs. (\ref{eq:intro02}) and (\ref{eq:intro04}), respectively. The extension of all these physical entropies from $3$ to $D, D > 3,$ dimensions is direct and then the parameter $\alpha$ of the involved orthogonal polynomials is directly proportional to $D$. The usefulness of high- and very high-dimensional quantum systems and phenomena has been amply shown in the the literature from general quantum mechanics and quantum field theory\cite{witten,yaffe1,herschbach,tsipis,chatterjee,avery,svidzinsky,dong,dehesa_2010,dehesa_2012,aptekarev} to quantum information \cite{bellomo, krenn,crann,irene}.\\

In this work we first study the asymptotic behavior of these integral functionals for large positive values of the parameter $\alpha$, while the other parameters are fixed. Then, as a separate case, we take $\mu=O(\alpha)$ as an additional large parameter. These integrals arise in the study of entropy-like functionals of R\'{e}nyi and Shannon types which describe various facets of the electronic spreading of the quantum probability density of the $D$-dimensional hydrogenic and harmonic systems in both position and momentum space with large and very large dimensionalities \cite{dehesa_2010,dehesa_2012,aptekarev}. These entropic measures are closely related to various fundamental and/or experimentally  accessible quantities (e.g., charge and momentum average densities, Thomas-Fermi and exchange potential energies, ...) of electronic systems (see e.g., \cite{dehesa_2012}). Moreover, they characterize  some uncertainty measures which have allowed to find the position-momentum  uncertainty relations of entropic type \cite{bialynicki3,zozor}. These relations are the mathematical formalizations of the uncertainty principle of quantum mechanics which generalize the Heisenberg uncertainty relation \cite{zozor2,toranzo}. 

The structure of this work is the following. First, in Section $2$ we give the basic asymptotics of Laguerre and Gegenbauer polynomials needed in the rest of the paper. Then, in Sections $3$ and $4$ we obtain the asymptotic expansions of the Laguerre and Gegenbauer integral functionals for large positive values of the parameter $\alpha$, while the other parameters are fixed. Finally, in Section $5$ we consider the asymptotic expansion of the Laguerre functionals for large positive values of the parameters $\alpha$ and $\mu$.

%\section{Well-known limits and further details of these limits}
\section{Basic asymptotics of Laguerre and Gegenbauer polynomials}
\label{sec:limits}
In this section we gather some well-known limits and we give some further details of these limits. We begin with the limits (\cite[Eqn.~(18.6.5)]{nist1})
\begin{equation}
\label{eq:limit01}
\lim_{\alpha\to \infty} \alpha^{-m}\mathcal{L}_{m}^{(\alpha)}(\alpha t) = \frac{(1-t)^{m}}{m!},
\end{equation}
and (\cite[Eqn.~(18.6.4)]{nist1})
\begin{equation}
\label{eq:limit02}
\lim_{\alpha \to \infty} \frac{1}{(2\alpha)_{m}}\mathcal{C}_{m}^{(\alpha)}(x) = \frac{x^{m}}{m!}.
\end{equation}
These relations can be used to obtain first approximations of the four integrals $I_{j}(m,\alpha)$ for large values of $\alpha$. Before showing more details, we give more information about these limits and we derive complete asymptotic expansions of $\mathcal{L}_{m}^{(\alpha)}(\alpha t) $  and $\mathcal{C}_{m}^{(\alpha)}(x) $ as $\alpha\to+\infty$.

We have the Taylor expansion
\begin{equation}
\label{eq:limit03}
\mathcal{L}_{m}^{(\alpha)}(x) =  \sum_{n=0}^m \frac{(x-\alpha)^{n}}{n!}\left.\bigl(\frac{d^n}{dx^n}\mathcal{L}_{m}^{(\alpha)}(x)\bigr)\right\vert_{x=\alpha}.
\end{equation}
We have the relation for the derivative (\cite[Eqn.~(18.9.23)]{nist1})
\begin{equation}
\label{eq:limit04}
\frac{d}{dx}\mathcal{L}_{m}^{(\alpha)}(x) =  -\mathcal{L}_{m-1}^{(\alpha+1)}(x),
\end{equation}
and this gives
\begin{equation}
\label{eq:limit05}
\mathcal{L}_{m}^{(\alpha)}(x) =  \sum_{n=0}^m (-1)^n\frac{(x-\alpha)^{n}}{n!}\mathcal{L}_{m-n}^{(\alpha+n)}(\alpha).
\end{equation}
We write this in the form
\begin{equation}
\label{eq:limit06}
\mathcal{L}_{m}^{(\alpha)}(\alpha t) =  \sum_{n=0}^m \frac{\alpha^{m-n}(1-t)^{m-n}}{(m-n)!}f_n(m;\alpha),\quad f_n(m;\alpha)=
\mathcal{L}_{n}^{(\alpha+m-n)}(\alpha),
\end{equation}
and we see that, because $f_0(m;\alpha)=1$, the term $n=0$ corresponds to the limit in \eqref{eq:limit01}.

A few other values of $f_n(m;\alpha)$ are
\begin{equation}
\begin{array}{@{}r@{\;}c@{\;}l@{}}
\label{eq:limit07}
f_1(m;\alpha)&=&m, \quad f_2=\frac12\bigl(m(m-1)-\alpha\bigr),\\ [8pt]
 f_3(m;\alpha)&=&\frac16\bigl(m(m-1)(m-2)+2\alpha-3m\alpha\bigr),\\[8pt]
 f_4(m;\alpha)&=& \frac{1}{24}\bigl(m(m-1)(m-2)(m-3)-2\alpha(3m^2-7m+3)+3\alpha^2\bigr).
\end{array}
\end{equation}
A recurrence relation for $f_n(m;\alpha)$ with respect to $n$ reads
\begin{equation}
\label{eq:limit08}
(n+1)f_{n+1}(m;\alpha)=(m-n)f_n(m;\alpha)-\alpha f_{n-1}(m;\alpha).
\end{equation}
This follows from the representations in terms of the Kummer function
(see (\cite[Eqn.~(18.11.2)]{nist1})
\begin{equation}
\label{eq:limit09}
\begin{array}{@{}r@{\;}c@{\;}l@{}}
f_n(m;\alpha)&=&\dsp{\binom{\alpha+m}{n}\ {}_1F_1(-n;\alpha+m+1-n;\alpha)}\\[8pt]
&=&\dsp{\binom{\alpha+m}{n}e^{\alpha}\ {}_1F_1(\alpha+m+1;\alpha+m+1-n;-\alpha)},
\end{array}
\end{equation}
and the recursion of  $\ {}_1F_1(a;b;z)$ in the second representation with respect to the $b$-direction  (see (\cite[Eqn.~(13.3.2)]{nist1}).

With mathematical induction, using \eqref{eq:limit08}, we conclude that
\begin{equation}
\label{eq:limit10}
f_{2n}(m;\alpha)=\bigO\left(\alpha^{n}\right), \quad f_{2n+1}(m;\alpha)=\bigO\left(\alpha^{n}\right),\quad \alpha\to \infty,
\end{equation}
and that the representation in \eqref{eq:limit06} has an asymptotic character for large $\alpha$, because of the decreasing order with respect to large $\alpha$ of pairs of successive terms.

We can rearrange the representation in \eqref{eq:limit06} into a polynomial of degree $m$ with decreasing powers of $\alpha$:
\begin{equation}
\label{eq:limit11}
\mathcal{L}_{m}^{(\alpha)}(\alpha t) =\frac{\alpha^{m}(1-t)^{m}}{m!} \sum_{n=0}^m \frac{g_n(m;t)}{\alpha^n},
\end{equation}
where the first coefficients are
\begin{equation}
\label{eq:limit12}
\begin{array}{@{}r@{\;}c@{\;}l@{}}
g_0(m,t)&=&1,\quad \dsp{g_1(m,t)=\frac{m(m+1-2mt)}{2(1-t)^2}},\\[8pt]
g_2(m,t)&=&\dsp{\frac{m(m-1)\left(3m^2(1-2t)^2-m(12t^2+8t-5)+16t+2\right)}{24(1-t)^4}.}
\end{array}
\end{equation}
However,  the form in \eqref{eq:limit06} with powers of $(1-t)$  is more convenient when using it to obtain asymptotic information of the integrals in \eqref{eq:intro01} and \eqref{eq:intro02}.

For the Gegenbauer polynomials a similar result is straightforward by using the explicit representation (see (\cite[Eqn.~(18.5.10)]{nist1})
\begin{equation}
\label{eq:limit13}
\mathcal{C}_{m}^{(\alpha)}(x) =\sum_{n=0}^{\left\lfloor m/2\right\rfloor}
\frac{(-1)^{n}\left(\alpha\right)_{m-n}}{n!\;(m-2n)!}
(2x)^{m-2n},
\end{equation}
and the term with $n=0$ corresponds to the limit in \eqref{eq:limit02}. In addition, successive terms are of lower order with respect to large values of $\alpha$. That is, denoting the terms by $T_n$, then one has $T_{n+1}/T_n=\bigO\left(\alpha^{-1}\right)$ as $\alpha \to\infty$.

\section{Asymptotic expansions of the Laguerre integrals}
%\section{Asymptotic expansions of the  integrals: the Laguerre case}
\label{sec:Lagcase}
In this section we obtain the asymptotic expansion of the R\'{e}nyi and Shannon-like integral functionals of Laguerre polynomials given by Eqs. (\ref{eq:intro01}) and (\ref{eq:intro02}), respectively, for large positive values of the parameter $\alpha$, while the other parameters are fixed. 
For the integral in \eqref{eq:intro01} we change the variable of integration by writing $x=\alpha t$, and obtain
\begin{equation}
\label{eq:Lag01}
I_{1} (m,\alpha)= \alpha^{\mu} \int\limits_{0}^{\infty}t^{\mu-1}e^{-\lambda \alpha t} \left|\mathcal{L}_{m}^{(\alpha)}(\alpha t)\right|^{\kappa}\,dt.
\end{equation}
For large values of $\alpha$, it follows from \eqref{eq:limit11} that
\begin{equation}
\label{eq:Lag02}
\mathcal{L}_{m}^{(\alpha)}(\alpha t)    \frac{\alpha^{m}(1-t)^{m}}{m!}\left(1+\bigO(1/\alpha)\right),
\end{equation}
and that $I_{1} (m,\alpha)\sim I_{1}^{(0)} (m,\alpha)$, where
\begin{equation}
\label{eq:Lag03}
 I_{1}^{(0)} (m,\alpha)= \alpha^{\mu}  \frac{\alpha^{\kappa m}}{(m!)^\kappa} \int\limits_{0}^{\infty}t^{\mu-1}e^{-\lambda \alpha t} \vert1-t\vert^{\kappa m}\,dt.
\end{equation}
An asymptotic expansion can be obtained by expanding 
\begin{equation}
\label{eq:Lag04}
(1-t)^{\kappa m}=\sum_{n=0}^\infty\frac{(-\kappa m)_n}{n!}t^n,
\end{equation}
 and invoking Watson's lemma (\cite[Chapter 3]{temme2}). This gives
\begin{equation}
\label{eq:Lag05}
 I_{1}^{(0)} (m,\alpha)\sim  \frac{\alpha^{\kappa m}\Gamma(\mu)}{\lambda^\mu (m!)^\kappa} \sum_{n=0}^{\infty}\frac{(\mu)_n(-\kappa m)_n}{n!(\alpha\lambda)^n}.
\end{equation}

When we use more terms of the representation in \eqref{eq:limit06}, we write
\begin{equation}
\label{eq:Lag06}
\mathcal{L}_{m}^{(\alpha)}(\alpha t) =   \frac{\alpha^{m}(1-t)^{m}}{m!}\sum_{n=0}^m \frac{m!}{(m-n)!\,\alpha^n}f_n(m;\alpha)(1-t)^{-n}.
\end{equation}
We expand
\begin{equation}
\label{eq:Lag07}
\left|\mathcal{L}_{m}^{(\alpha)}(\alpha t)\right|^{\kappa}\sim
\frac{\alpha^{\kappa m}}{(m!)^\kappa} (1-t)^{\kappa m}\sum_{j=0}^\infty \frac{A_{j}}{\alpha^j}(1-t)^{-j},\quad 0\le t < 1,
\end{equation}
where the first coefficients are 
\begin{equation}
\begin{array}{@{}r@{\;}c@{\;}l@{}}
\label{eq:Lag08}
A_{0}&=&1,\quad A_1=\kappa mf_1(m;\alpha), \\ [8pt]
A_{2}&=&\frac12\kappa m\bigl(2mf_2(m;\alpha)-2f_2(m;\alpha)-mf_1(m;\alpha)^2+\kappa mf_1(m;\alpha)^2\bigr). 
%A_{3}&=&\dsp{\frac{\kappa m(6f_3m^2-18f_3m+12f_3-6f_1m^2f_2+6f_1mf_2+2f_1^3m^2+6\kappa f_1m^2f_2-6\kappa f_1mf_2-3\kappa f_1^3m^2+\kappa^2f_1^3m^2)}{6\alpha^3}}.
\end{array}
\end{equation}

The representation in \eqref{eq:Lag06} has an asymptotic character for large $\alpha$ starting with the first term equal to 1, and, although the coefficients $A_{j}$  depend on $\alpha$, the series in \eqref{eq:Lag07} has an asymptotic character as well. We obtain
\begin{equation}
\label{eq:Lag09}
 I_{1}(m,\alpha)\sim  \sum_{j=0}^{\infty}\frac{A_{j} }{\alpha^j}I_{1}^{(j)} (m,\alpha),
 \end{equation}
where
\begin{equation}
\label{eq:Lag10}
 I_{1}^{(j)} (m,\alpha) =\alpha^{\mu}  \frac{\alpha^{\kappa m}}{(m!)^\kappa}\int_0^\infty t^{\mu-1}e^{-\lambda \alpha t}(1-t)^{\kappa m-j}\,dt.
\end{equation}

\begin{remark}\label{rem:rem01}{\ }

\begin{enumerate}
\item
It should be observed that the expansion in \eqref{eq:Lag07} contains negative powers of $(1-t)$. This gives divergent integrals in  \eqref{eq:Lag10} 
when $\kappa m-j\le -1$. For the asymptotic results  this is not relevant, because in the application of Watson's lemma we can concentrate on small intervals $[0,t_0]$, $t_0\in(0,1)$, of the Laplace integrals, even in the starting integral in \eqref{eq:Lag01}.
\item
It follows  from \eqref{eq:Lag02} that for small values of $t$ and large $\alpha$ the function 
$\mathcal{L}_{m}^{(\alpha)}(\alpha t)$ is positive. Hence, to obtain the expansion in \eqref{eq:Lag07} we may skip the absolute values.  
\eoremark
\end{enumerate}

\end{remark}

With this in mind, we can expand the functions defined in \eqref{eq:Lag10}  in the form
\begin{equation}
\label{eq:Lag11}
 I_{1}^{(j)} (m,\alpha) \sim   \frac{\alpha^{\kappa m}\Gamma(\mu)}{\lambda^\mu (m!)^\kappa} \sum_{n=0}^{\infty}\frac{B_{j,n}}{\alpha^n},\quad B_{j,n}=\frac{(\mu)_n(j-\kappa m)_n}{n!\,\lambda^n},
\end{equation}
and when using this in \eqref{eq:Lag09} we obtain
\begin{equation}
\label{eq:Lag12}
 I_{1}(m,\alpha)\sim   \frac{\alpha^{\kappa m}\Gamma(\mu)}{\lambda^\mu (m!)^\kappa} \sum_{k=0}^{\infty}\frac{C_{k}(\alpha) }{\alpha^k},
 \quad C_k(\alpha)=\sum_{j=0}^kA_jB_{j, k-j}.
 \end{equation}
In this expansion the coefficients $C_k(\alpha)$ are in terms of $f_j(m;\alpha)$, which are polynomials of $\alpha$; see \eqref{eq:limit07},  \eqref{eq:limit10} and \eqref{eq:Lag08}. 
Considering the above construction of $C_k(\alpha)$ and the behavior of $A_j$ defined in \eqref{eq:Lag07}, we conclude that, as in  \eqref{eq:limit10} for $f_n(m,\alpha)$, 
\begin{equation}
\label{eq:Lag13}
C_{2k}(\alpha)=\bigO\left(\alpha^{k}\right), \quad C_{2k+1}(\alpha)=\bigO\left(\alpha^{k}\right),\quad \alpha\to \infty.
\end{equation}
As a consequence, we can rearrange the expansion in \eqref{eq:Lag12} to obtain an expansion in negative powers of $\alpha$. 

To verify this, we split up
\begin{equation}
\label{eq:Lag14}
 \sum_{k=0}^{\infty}\frac{C_{k} (\alpha)}{\alpha^k}= \sum_{k=0}^{\infty}\frac{C_{2k} (\alpha)}{\alpha^{2k}}+ \sum_{k=0}^{\infty}\frac{C_{2k+1}(\alpha) }{\alpha^{2k+1}},
 \end{equation}
and we can write
\begin{equation}
\label{eq:Lag15}
 \sum_{k=0}^{\infty}\frac{C_{k}(\alpha) }{\alpha^k}= \sum_{k=0}^{\infty}\frac{C^{(0)}_k(\alpha)}{\alpha^{k}}+ 
\frac{1}{\alpha}  \sum_{k=0}^{\infty}\frac{C^{(1)}_k(\alpha)}{\alpha^{k}},
 \end{equation}
where $C^{(0)}_k(\alpha)$ and $C^{(1)}_k(\alpha)$ have expansions in negative powers of $\alpha$:
\begin{equation}
\label{eq:Lag16}
C^{(0)}_k(\alpha)\sim \sum_{k=0}^{\infty}\frac{C_{j,2k} }{\alpha^k}, \quad C^{(1)}_k(\alpha)\sim \sum_{k=0}^{\infty}\frac{C_{j,2k+1} }{\alpha^k}.
\end{equation}

In this way,  we can obtain an expansion in negative powers of $\alpha$:
\begin{equation}
\label{eq:Lag17}
 \sum_{k=0}^{\infty}\frac{C_{k}(\alpha) }{\alpha^k}= \sum_{k=0}^{\infty}\frac{D_{k} }{\alpha^k},\quad D_k=\sum_{j=0}^k C_{j,2k-2j}+\sum_{j=0}^{k-1}C_{j,2k-1-2j}
 \end{equation}
and the first $D_k$ are
\begin{equation}
\label{eq:Lag18}
\begin{array}{@{}r@{\;}c@{\;}l@{}}
D_1&=&C_{0,0}=1,\quad D_1=C_{0,2}+C_{1,0}+C_{0,1},\\[8pt]
D_2&=&C_{0,4}+C_{1,2}+C_{2,0}+C_{0,3}+C_{1,1}.
\end{array}
\end{equation}
All coefficients can be computed by straightforward manipulations of series with the help of software for symbolic computations.\footnote{ Maple codes for the determination of the coefficients $D_k$ and of the coefficients appearing in other asymptotic expansions given in this paper are available from the authors upon request.}

We summarize the above results as follows.
\begin{proposition}\label{prop:P1}
Let $\alpha, \lambda, \kappa$, and $\mu$ be positive real numbers, and $m$ a positive natural number. Then, for the R\'{e}nyi-like integral
\begin{equation}
\label{eq:Lag19}
I_{1} (m,\alpha)= \int\limits_{0}^{\infty}x^{\mu-1}e^{-\lambda x} \left|\mathcal{L}_{m}^{(\alpha)}(x)\right|^{\kappa}\,dx, 
\end{equation}
we have the asymptotic expansion 
\begin{equation}
\label{eq:Lag20}
 I_{1}(m,\alpha)\sim   \frac{\alpha^{\kappa m}\Gamma(\mu)}{\lambda^\mu (m!)^\kappa} \sum_{k=0}^{\infty}\frac{D_{k} }{\alpha^k},\quad \alpha \to\infty. 
 \end{equation}
The first coefficients are
\begin{equation}
\label{eq:Lag21}
D_{0}=1,\quad D_1=\frac{\kappa m (-2\mu+m\lambda+\lambda)}{2\lambda},
\end{equation}
and
\begin{equation}
\label{eq:Lag22}
\begin{array}{@{}r@{\;}c@{\;}l@{}}
D_2&=&
\kappa m\bigl(-12\mu \lambda \kappa m^2+24\mu \lambda -12\mu \lambda \kappa m-4m^2\lambda ^2-6m\lambda ^2+
3m^3\lambda ^2\kappa \\ [8pt]
&&-12\mu ^2+12\mu ^2\kappa m-12\mu +12\mu \kappa m+6\lambda ^2\kappa m^2-2\lambda ^2+3\lambda ^2\kappa m\bigr)/(24\lambda ^2).
\end{array}
\end{equation}

\end{proposition}

\begin{remark}
\label{rem:rem02}
To obtain a result for the Shannon-like integrals $I_2(m,\alpha)$ defined in \eqref{eq:intro02}, we differentiate the expansion in \eqref{eq:Lag20} with respect to $\kappa$, and take $\kappa=2$ afterwards. We have
 \begin{equation}
\label{eq:Lag23}
I_2(m,\alpha)=2\left.\frac{\partial}{\partial \kappa}I_1(m,\alpha)\right\vert_{\kappa=2}\sim \frac{\alpha^{2 m}\Gamma(\mu)}{\lambda^\mu (m!)^2} 
\bigl(\log\frac{\alpha^{2m}}{(m!)^2}\sum_{k=0}^{\infty}\frac{D_{k} }{\alpha^k}+2\sum_{k=0}^{\infty}\frac{D_{k}^\prime }{\alpha^k}\bigr),
\end{equation}
for $\alpha \to\infty$ and the rest of parameters are fixed. The derivatives in $D_k$ are with respect to $\kappa$. 
\eoremark
\end{remark}

\begin{example}\label{ex:E1}
We have the special case for $\kappa=2$ and $\lambda=1$ (see \cite[Page~478]{Prudnikov})
\begin{equation}
\label{eq:Lag24}
\begin{array}{@{}r@{\;}c@{\;}l@{}}
I_{1} (m,\alpha)&=&\dsp{ \int\limits_{0}^{\infty}x^{\mu-1}e^{-x} \mathcal{L}_{m}^{(\alpha)}(x)^{2}\,dx=\frac{(\alpha+1)_m(\alpha+1-\mu)_m\Gamma(\mu)}{m!\,m!}\ \times}\\ [8pt]
&&\quad\quad\dsp{\ {}_3F_2\left(-m,\mu,\mu-\alpha;\alpha+1,\mu-\alpha-m;1\right).}
\end{array}
\end{equation}
We can expand the finite $\ {}_3F_2$-function term by term in negative powers of $\alpha$, and the Pochhammer symbols can be expanded as well:
\begin{equation}
\label{eq:Lag25}
\begin{array}{@{}r@{\;}c@{\;}l@{}}
&&(\alpha+1)_m=\dsp{\frac{\Gamma(\alpha+1+m)}{\Gamma(\alpha+1)}\sim
\alpha^m\left(1+\frac{m(m+1)}{2\alpha}+\ldots\right),}\\[8pt]
&&(\alpha+1-\mu)_m = \dsp{\frac{\Gamma(\alpha+1-\mu+m)}{\Gamma(\alpha+1-\mu)}\sim
\alpha^m\left(1+\frac{m(m+1-\mu)}{2\alpha}+\ldots\right).}
\end{array}
\end{equation}
More terms follow from \cite[\S6.5.1]{temme2}. We obtain the expansion as  in \eqref{eq:Lag20}, with coefficients $D_0=1$, $D_1=m(1+m-2\mu)$ and 
\begin{equation}
\label{eq:Lag26}
D_{2}= \tfrac16 m \left(-1+6 \mu-6 \mu^2+12 \mu^2 m-12 m^2 \mu+4 m^2+3 m^3\right),
\end{equation}
which confirms those for \eqref{eq:Lag20} when $\kappa=2$ and $\lambda=1$.
\eoexample
\end{example}

\section{Asymptotic expansions of the Gegenbauer integrals}
\label{sec:gegcase}
In this section we obtain the asymptotic expansion of the R\'{e}nyi and Shannon-like integral functionals of Gegenbauer polynomials given by Eqs. (\ref{eq:intro03}) and (\ref{eq:intro04}), respectively, for large positive values of the parameter $\alpha$, while the other parameters are fixed. For the integral in \eqref{eq:intro03} we need to consider separately the following two cases: $c\ne d$ and $c=d$. 
\subsection{The case $c\ne d$}\label{sec:dc}
We assume that $c<d$, and we observe that $c>d$ follows from interchanging $a$ and $b$ and $c$ and $d$. 

The limit in \eqref{eq:limit02} can be written in the equivalent form
\begin{equation}
\label{eq:Geg01}
\lim_{\alpha \to \infty} \frac{1}{(\alpha)_{m}}\mathcal{C}_{m}^{(\alpha)}(x) = \frac{(2x)^{m}}{m!},
\end{equation}
and we write  the representation given in \eqref{eq:limit13} in the form
\begin{equation}
\label{eq:Geg02}
\begin{array}{@{}r@{\;}c@{\;}l@{}}
\mathcal{C}_{m}^{(\alpha)}(x) &=&\dsp{\frac{(2x)^m(\alpha)_m}{m!}
\sum_{n=0}^{\left\lfloor m/2\right\rfloor}
\frac{(-1)^{n}m!\left(\alpha\right)_{m-n}}{(\alpha)_m n!\;(m-2n)!}(2x)^{-2n}}\\[8pt]
&=&\dsp{\frac{(2x)^m(\alpha)_m}{m!}\sum_{n=0}^{\left\lfloor m/2\right\rfloor}
\frac{f_n(m;\alpha)}{\alpha^n}x^{-2n}}\\[8pt]
&=&\dsp{\frac{(2x)^m(\alpha)_m}{m!} \left(1-\frac{m(m-1)}{4x^2(\alpha+m-1)}+\bigO\left(\alpha^{-2}\right)\right).}
\end{array}
\end{equation}
Because successive terms are of lower order of $\alpha$, this is an asymptotic representation, if $x\ne0$. Observe that $f_n(m;\alpha)=\bigO(1)$ as $\alpha\to\infty$.

We expand
\begin{equation}
\label{eq:Geg03}
\left|\mathcal{C}^{(\alpha)}_{m}(x)\right|^{\kappa}=\frac{(2x)^{\kappa m}\bigl((\alpha)_m\bigr)^\kappa}{(m!)^\kappa}\left\vert\sum_{j=0}^\infty\frac{A_j}{\alpha^j}x^{-2j}\right\vert
\end{equation}
where the first coefficients are
\begin{equation}
\begin{array}{@{}r@{\;}c@{\;}l@{}}
\label{eq:Geg04}
A_{0}&=&1,\quad A_1=\kappa f_1(m;\alpha), \\ [8pt]
A_{2}&=&\frac12\kappa \bigl(2f_2(m;\alpha)-f_1(m;\alpha)^2+\kappa f_1(m;\alpha)^2\bigr). 
%A_{3}&=&\dsp{\frac{\kappa m(6f_3m^2-18f_3m+12f_3-6f_1m^2f_2+6f_1mf_2+2f_1^3m^2+6\kappa f_1m^2f_2-6\kappa f_1mf_2-3\kappa f_1^3m^2+\kappa^2f_1^3m^2)}{6\alpha^3}}.
\end{array}
\end{equation}
This gives for the integral in \eqref{eq:intro03} the expansion
\begin{equation}
\label{eq:Geg05}
I_{3}(m,\alpha)  \sim \frac{2^{\kappa m}\bigl((\alpha)_m\bigr)^\kappa}{(m!)^\kappa} \sum_{j=0}^\infty\frac{A_j}{\alpha^j}I_{3}^{(2j)}(m,\alpha),
\end{equation}
where\footnote{We refer to Remark \ref{rem:rem01} for an interpretation of the possibly divergent integrals.}
\begin{equation}
\label{eq:Geg06}
I_{3}^{(2j)}(m,\alpha)  = \int_{-1}^{1} (1-x)^{c\alpha+a} (1+x)^{d\alpha+b}\left|x\right|^{\kappa m-{2j}}\,dx, \quad j=0,1,2,\ldots
\end{equation}

We write this in the form 

\begin{equation}
\label{eq:Geg07}
I_{3}^{(2j)}(m,\alpha)  = \int_{-1}^{1} (1-x)^{a} (1+x)^{b}\left|x\right|^{\kappa m-{2j}}e^{-\alpha\phi(x)}\,dx, 
\end{equation}
where
\begin{equation}
\label{eq:Geg08}
\phi(x)=-c\log(1-x)-d\log(1+x).
\end{equation}
This function assumes a minimum at the internal point $x_m=(d-c)/(d+c)$, the saddle point, and we can apply Laplace's method (see, for example,  \cite[Chapter 3]{temme2})  to obtain an asymptotic representation.

When $c<d$ we have $x_m\in(0,1)$ and the contribution of the interval $(-1,0)$ is exponentially small compared with that of $(0,1)$.
Hence, we replace the interval $(-1,1)$ by $(0,1)$ and introduce the new variable of integration $y$ by writing
\begin{equation}
\label{eq:Geg09}
\phi(x)-\phi(x_m)=\tfrac12 y^2,\quad \sign(y)=\sign(x-x_m).
\end{equation}
In addition we extend the $y$-interval into $(-\infty,\infty)$. We obtain
\begin{equation}
\label{eq:Geg10}
I_{3}^{(2j)}(m,\alpha)  \sim e^{-\alpha \phi(x_m)}  \int_{-\infty}^{\infty} e^{-\frac12\alpha y^2} f_j(y)\,dy, 
\end{equation}
where
\begin{equation}
\label{eq:Geg11}
f_j(y)=(1-x)^{a} (1+x)^{b}x^{\kappa m-{2j}}\frac{dx}{dy}.
\end{equation}
With the expansion $\dsp{f_j(y)=\sum_{k=0}^{\infty}c_k^{(2j)}y^k}$ the asymptotic result follows: 
\begin{equation}
\label{eq:Geg12}
I_{3}^{(2j)}(m,\alpha)  \sim e^{-\alpha \phi(x_m)}\sqrt{\frac{2\pi}{\alpha}}\sum_{k=0}^{\infty} c_{2k}^{(2j)}\frac{2^k\left(\frac12\right)_k}{\alpha^k},\quad \alpha\to\infty,
\end{equation}
where
\begin{equation}
\label{eq:Geg13}
\phi(x_m)=-c\log\frac{2c}{c+d}-d\log\frac{2d}{c+d}.
\end{equation}

To evaluate the coefficients $c_k^{(2j)}$ we derive from  \eqref{eq:Geg09} those in the expansion $\dsp{x=x_m+\sum_{k=1}^\infty a_k y^k}$.  We have
\begin{equation}
\label{eq:Geg14}
a_1=2\sqrt{\frac{cd}{(c+d)^3}},\quad a_2=\frac{2(c-d)}{3(c+d)^2}, \quad a_3=\frac{c^2-11cd+d^2}{9a_1(c+d)^4},
 \end{equation}
and next
\begin{equation}
\label{eq:Geg15}
\begin{array}{@{}r@{\;}c@{\;}l@{}}
c_0^{(2j)}&=&\dsp{a_1\left(\frac{2c}{c+d}\right)^{a}\left(\frac{2d}{c+d}\right)^{b}\left(\frac{d-c}{c+d}\right)^{\kappa m-2j},}\\[8pt]
c_1^{(2j)}&=&\dsp{c_0^{(2j)} a_1\frac{(c+d)\bigl(6cd(\kappa m-2j)+(d-c)(3bc-3ad+2c-2d)\bigr)}{6cd(d-c)}}.
\end{array}
 \end{equation}

Using the expansions of \eqref{eq:Geg12} in \eqref{eq:Geg05} we obtain 
\begin{equation}
\label{eq:Geg16}
I_{3}(m,\alpha)  \sim e^{-\alpha \phi(x_m)}\sqrt{\frac{2\pi}{\alpha}}\frac{2^{\kappa m}\bigl((\alpha)_m\bigr)^\kappa}{(m!)^\kappa} \sum_{k=0}^\infty\frac{C_k(\alpha)}{\alpha^k},
\end{equation}
with first coefficients
\begin{equation}
\label{eq:Geg17}
\begin{array}{@{}r@{\;}c@{\;}l@{}}
C_0(\alpha)&=&A_0c_0^{(0)},\quad C_1(\alpha)=A_0c_2^{(0)}+A_1c_0^{(2)},\\[8pt]
C_2(\alpha)&=&3A_0c_4^{(0)}+A_1c_2^{(2)}+A_2c_0^{(4)}.
\end{array}
 \end{equation}
These coefficients are $\bigO(1)$ as $\alpha\to\infty$. As explained for Proposition~\ref{prop:P1}, we can expand them, rearrange the series in \eqref{eq:Geg16}, and obtain an expansion in negative powers of~$\alpha$. 

We summarize the above results as follows.

\begin{proposition}\label{prop:P2}
Let $a, b, c, d$, and $\kappa$ be positive real numbers, $c<d$, and $m$ a positive natural number. Then, for the R\'{e}nyi-like integral
\begin{equation}
\label{eq:Geg18}
I_{3} (m,\alpha)= \int_{-1}^{1} (1-x)^{c\alpha+a} (1+x)^{d\alpha+b}\left|\mathcal{C}^{(\alpha)}_{m}(x)\right|^{\kappa}\,dx,  
\end{equation}
we have the asymptotic expansion given in \eqref{eq:Geg16}, which can be converted into the form
\begin{equation}
\label{eq:Geg19}
I_{3}(m,\alpha)  \sim e^{-\alpha \phi(x_m)}\sqrt{\frac{2\pi}{\alpha}}\frac{2^{\kappa m}\bigl((\alpha)_m\bigr)^\kappa}{(m!)^\kappa} \sum_{k=0}^\infty\frac{D_k}{\alpha^k},\quad \alpha\to\infty,
\end{equation}
with coefficients $D_k$ not depending on $\alpha$. The first coefficient is
\begin{equation}
\label{eq:Geg20}
D_0=C_0=a_1 \left(\frac{2c}{c+d}\right)^{a} \left(\frac{2d}{c+d}\right)^{b} \left(\frac{d-c}{c+d}\right)^{\kappa m},
\end{equation}
where $a_1$ is given in \eqref{eq:Geg14}.
\end{proposition}

\begin{remark}\label{rem:rem03}
A result for the Shannon-like integral $I_4(m,\alpha)$ defined in \eqref{eq:intro04} follows from differentiating the expansion in \eqref{eq:Geg19} with respect to $\kappa$, and taking $\kappa=2$ afterwards. We have
  \begin{equation}
\label{eq:Geg21}
\begin{array}{@{}r@{\;}c@{\;}l@{}}
&&I_4(m,\alpha)=\dsp{2\left.\frac{\partial}{\partial \kappa}I_3(m,\alpha)\right\vert_{\kappa=2}\ \sim} \\[8pt]
&&\quad\quad
\dsp{ e^{-\alpha\phi(x_m)}\sqrt{\frac{2\pi}{\alpha}} 
\frac{2^{2 m}\bigl((\alpha)_m\bigr)^2}{(m!)^2} 
\left(\log\frac{2^{2m}\bigl((\alpha)_m\bigr)^2}{(m!)^2} 
\sum_{k=0}^{\infty}\frac{D_{k} }{\alpha^k}+2\sum_{k=0}^{\infty}\frac{D_{k}^\prime }{\alpha^k}\right),}
\end{array}
\end{equation}
for $\alpha \to\infty$ and the rest of parameters are fixed. Here again, the derivatives in $D_k$ are with respect to $\kappa$.  For the corresponding logarithmic case when $c=d$ we refer to Remark~\ref{rem:rem04}. The special form of $I_4(m,\alpha)$ in \eqref{eq:intro04} with $a=b=-\frac12$ and $c=d=1$ does not follow from \eqref{eq:Geg21}. For the case $c=d=1$ we refer to \S\ref{sec:disc}.

\eoremark
\end{remark}

\begin{example}\label{ex:E2}
We have the special case (see \cite[Eqn. 6, Page~562]{Prudnikov}) for $\kappa=2$, $a=-\frac12$, $b=2m-\frac32$, $c=1$, $d=3$, 
\begin{equation}
\label{eq:Geg22}
\begin{array}{@{}r@{\;}c@{\;}l@{}}
I_{3}(m,\alpha) & = &\dsp{ \int_{-1}^{1} (1-x)^{\alpha-\frac12} (1+x)^{3\alpha+2m-\frac32}\bigl(\mathcal{C}^{(\alpha)}_{m}(x)\bigr)^{2}\,dx}\\[8pt]
&=&\dsp{\frac{\sqrt{\pi}(2\alpha)_{2m}}{\left(2^{m}\left(\alpha+\frac12 \right)_{m}\,m!\right)^2}
 \frac{\Gamma\left(\alpha+2m+\frac12\right)\Gamma\left(3\alpha+2m-\frac12\right)}{\Gamma\left(2\alpha\right)\Gamma\left(2\alpha+2m+\frac12\right)}.     }
\end{array}
\end{equation}
Expanding this for large $\alpha$, we obtain the first-order approximation
\begin{equation}
\label{eq:Geg23}
I_3\sim \sqrt{\frac{\pi}{\alpha}}\,\frac{3^{3\alpha+2m-1}\alpha^{2m}}{2^{4\alpha+2m}\left(m!\right)^2}.
\end{equation}
The same result follows from \eqref{eq:Geg19} with the first term $D_0$ and the special choice of the parameters.
\eoexample
\end{example}

\subsection{The case $c = d=1$}\label{sec:disc}
In this case we write \eqref{eq:intro03} in the form
\begin{equation}
\label{eq:Geg24}
I_{3}(m,\alpha)  = \int_{-1}^{1} (1-x)^{a} (1+x)^{b}e^{-\alpha\phi(x)}\left|\mathcal{C}^{(\alpha)}_{m}(x)\right|^{\kappa}\,dx, 
\end{equation}
where $\phi(x)$ is defined in \eqref{eq:Geg08}. It is symmetric on $(-1,1)$ and has a saddle point $x_m$ at $x=0$, with $\phi(0)=0$. We use the transformation given in \eqref{eq:Geg09}, and obtain
\begin{equation}
\label{eq:Geg25}
I_{3}(m,\alpha)  = \int_{-\infty}^{\infty} f(y)e^{-\frac12\alpha y^2}\left|\mathcal{C}^{(\alpha)}_{m}(x)\right|^{\kappa}\,dy, \quad f(y)=(1-x)^{a} (1+x)^{b}\frac{dx}{dy}.
\end{equation}

In this case, with the saddle point at the origin, we cannot use the relation that follows from the limit given in \eqref{eq:Geg01}, that is, 
$\dsp{\mathcal{C}^{(\alpha)}_{m}(x)\sim \frac{(2x)^m}{m!}(\alpha)_m}$. This relation is useless in a small interval around the origin; it does not hold uniformly with respect to small values of $x$. Instead, we may consider an expansion in ascending powers of $x$, for example in the notation of the Gauss hypergeometric function,
\begin{equation}
\label{eq:Geg26}
\begin{array}{@{}r@{\;}c@{\;}l@{}}
\mathcal{C}^{(\alpha)}_{2m}(x)& = &\dsp{(-1)^m\frac{(\alpha)_m}{m!}\ {}_2F_1\left(-m;m+\alpha;\tfrac12;x^2\right), }\\[8pt]
\mathcal{C}^{(\alpha)}_{2m+1}(x)& = &\dsp{(-1)^m\frac{(\alpha)_{m+1}}{m!}2x\ {}_2F_1\left(-m;m+\alpha+1;\tfrac32;x^2\right), }
\end{array}
\end{equation}
which are rearrangements of the representation in the first line of \eqref{eq:Geg02}. These forms clearly show that the expansions in powers of $x$ do not give asymptotic representations for large $\alpha$, unless $x=o(1/\sqrt{\alpha})$. Because $x\sim y/\sqrt{2}$ (see coefficient $a_1$ in \eqref{eq:Geg14}), an expansion of $\left|\mathcal{C}^{(\alpha)}_{m}(x)\right|^\kappa$ in powers of $y$ is useless for obtaining the large $\alpha$ expansion of the integral given in  \eqref{eq:Geg25}.\footnote{See also the discussion in \cite[\S5.1]{Bruijn} about the range of a saddle point.}

From the literature (see, for example, \cite[\S24.2]{temme2} we know that large $\alpha$ approximations of $\mathcal{C}^{(\alpha)}_{m}(x)$ can be given in terms of Hermite polynomials, and these are uniformly valid in an $x$-interval around the origin. We also know the simple 
relation (see \cite[Eqn.~(18.7.24)]{nist1}
\begin{equation}
\label{eq:Geg27}
\lim_{\alpha\to\infty}\alpha^{-\frac12m}\mathcal{C}^{(\alpha)}_{m}\left(\alpha^{-\frac12}\,x\right)=\frac{H_m(x)}{m!}.
\end{equation}
When we use this as a first order asymptotic relation in \eqref{eq:Geg25} and observe that  $x=y/\sqrt{2}+\bigO\left(y^3\right)$, and replace $f(y)$ by $f(0)=1/\sqrt{2}$, we obtain the following asymptotics of the R\'{e}nyi-like integral
\begin{equation}
\label{eq:Geg28}
I_{3}(m,\alpha)  \sim \frac{\alpha^{\frac12\kappa m}}{\sqrt{2}\,(m!)^\kappa} \int_{-\infty}^{\infty} e^{-\frac12\alpha y^2}\left|H_{m}\left(y\sqrt{\alpha/2}\right)\right|^{\kappa}\,dy,
\end{equation}
for $\alpha \to\infty$ and the rest of parameters are fixed.
Using the orthogonality relation of the Hermite polynomials, we can evaluate the integral when $\kappa=2$  and find
\begin{equation}
\label{eq:Geg29}
I_{3}(m,\alpha)  \sim \sqrt{\frac{\pi}{\alpha}}\frac{(2\alpha)^m}{m!},\quad \alpha\to\infty.
\end{equation}

This result is not very detailed; for example, it does not show the parameters $a$ and $b$. In fact we can try an expansion of the form
\begin{equation}
\label{eq:Geg30}
\begin{array}{@{}r@{\;}c@{\;}l@{}}
I_{3}(m,\alpha)  &\sim&\dsp{ \sqrt{\frac{\pi}{\alpha}}\frac{(2\alpha)^m}{m!}\sum_{k=0}^\infty\frac{D_k}{\alpha^k},\quad D_0=1,}\\ [8pt]
D_1&=&\frac18\left(2(2m+1) \bigl( (a-b)^2-(a+b)\bigr)+2 m^2-14m-3\right).
\end{array}
 \end{equation}
 In  \S\ref{sec:gegher} we explain  how to obtain $D_1$ and more coefficients.

\begin{example}\label{ex:E3}
For $\kappa=2$, $a=-\frac12$,  $b=-\frac32$ , $c=d=1$, the corrected result of  \cite[Eqn. 7, Page~562]{Prudnikov} is
\begin{equation}
\label{eq:Geg31}
I_{3}(m,\alpha) = \int_{-1}^{1} (1-x)^{\alpha-\frac12} (1+x)^{\alpha-\frac32}\bigl(\mathcal{C}^{(\alpha)}_{m}(x)\bigr)^{2}\,dx=\frac{\sqrt\pi\,(2\alpha)_{m}}{m!}
 \frac{\Gamma\left(\alpha-\frac12\right)}{\Gamma\left(\alpha\right)}.
\end{equation}
When we expand the right-hand side for large values of $\alpha$, we obtain 
\begin{equation}
\label{eq:Geg32}
I_{3}(m,\alpha)  \sim \sqrt{\frac{\pi}{\alpha}}\frac{(2\alpha)^m}{m!}\Bigl(1+
\frac{2m^2-2m+3}{8\alpha}\Bigr),
 \end{equation}
which corresponds to the estimate given in \eqref{eq:Geg30} when we take $a=-\frac12$ and  $b=-\frac32$.
It is easy to verify that this correspondence does not happen when we use the asymptotic relation that follows from  \eqref{eq:Geg01} instead of the one that follows from  \eqref{eq:Geg27}.
\eoexample
\end{example}

\begin{remark}\label{rem:rem04}
A result for the Shannon-like integral $I_4(m,\alpha)$ defined in \eqref{eq:intro04} follows from differentiating \eqref{eq:Geg28} with respect to $\kappa$ and putting $\kappa=2$ afterwards. It seems not to be possible to give a large-$\alpha$ expansion of  the resulting integral.
\eoremark
\end{remark}

\subsection{Hermite-type expansion of the Gegenbauer polynomials}\label{sec:gegher}
Here we find more asymptotic details of the approximation in \eqref{eq:Geg30}. To do that we expand the Gegenbauer polynomials in an asymptotic representation in terms of the Hermite polynomials of the form
\begin{equation}
\label{eq:Geg33}
\mathcal{C}^{(\alpha)}_{m}(z)\sim\frac{\alpha^{\frac12m}}{m!}\left(H_m\left(z\sqrt{\alpha}\right)\sum_{k=0}^\infty\frac{c_k}{\alpha^k}+\frac{m}{\sqrt{\alpha}}
H_{m-1}\left(z\sqrt{\alpha}\right)\sum_{k=0}^\infty\frac{d_k}{\alpha^k}\right).
\end{equation}
The first coefficients are
\begin{equation}
\label{eq:Geg34}
\begin{array}{@{}r@{\;}c@{\;}l@{}}
c_0&=&1,\quad d_0=0, \\ [8pt]
c_1&=&-\alpha z^2 m,\quad d_1=\frac13\alpha z\left(4 z^2+6 \alpha z^2-4 z^2 m-3+3 m\right).
\end{array}
\end{equation}

This expansion is valid for large $\alpha$ and bounded  $m$ and $z\sqrt{\alpha}$. The coefficients $c_k$ and $d_k$ depend on $\alpha$. After rearranging the expansion and putting $z=x/\sqrt{\alpha}$ we find
\begin{equation}
\label{eq:Geg35}
\mathcal{C}^{(\alpha)}_{m}\left(\alpha^{-\frac12}\,x\right)\sim\frac{\alpha^{\frac12m}}{m!}\left(H_m(x)\sum_{k=0}^\infty \frac{p_k}{\alpha^k}+\frac{m}{\alpha}H_{m-1}(x)\sum_{k=0}^\infty \frac{q_k}{\alpha^k}\right).
\end{equation}
The coefficients do not depend on $\alpha$ and the first few are
\begin{equation}
\label{eq:Geg36}
\begin{array}{@{}r@{\;}c@{\;}l@{}}
p_0&=&1,\quad q_0=\tfrac14x\left(2x^2+2m-1\right),\quad p_1=\tfrac18m(m-2x^2-2), \\ [8pt]
q_1&=&\tfrac1{192} x\Bigl(3+24m-42m^2+12m^3+\left(400m-48m^2-640\right)x^2\ +\\ [8pt]
&&\quad\quad\left(1280-384m\right)x^4\Bigr).
\end{array}
\end{equation}
This expansion is valid for large $\alpha$ and bounded $x$ and $m$. 

For uniform expansions in which $\alpha$ and $m$ may be of the same order, we refer to \cite[\S24.2]{temme2}. The simpler asymptotic results given above are not available in the literature, and we show how to find the coefficients.

 We start with the representation
\begin{equation}
\label{eq:Geg37}
\mathcal{C}^{(\alpha)}_{m}(z)=\frac{1}{2\pi i}\int_\calC\frac{1}{(1-2zt+t^2)^\alpha}\,\frac{dt}{t^{m+1}},
\end{equation}
where $\calC$ is a circle with radius smaller than 1. We write this the form
\begin{equation}
\label{eq:Geg38}
\begin{array}{@{}r@{\;}c@{\;}l@{}}
\mathcal{C}^{(\alpha)}_{m}(z)&=&\dsp{\frac{1}{2\pi i}\int_\calC e^{-\alpha\left(t^2-2zt\right)}h(t)\,\frac{dt}{t^{m+1}},}\\[8pt]
h(t)&=&\dsp{e^{-\alpha\left(\log\left(1-2zt+t^2\right)-t^2+2zt\right)}.}
\end{array}
\end{equation}

When we expand $h(t)$ in powers of $t$, we obtain a simple finite expansion in which each successive term contains a  Hermite polynomial of lower degree. A slightly different approach is given in \cite{Lopez}. In the problem to find more details of the expansion in \eqref{eq:Geg33}, it is more convenient to use an expansion with only two Hermite polynomials.

When we replace $h(t)$ by its value at the origin, $h(0)=1$, we obtain
\begin{equation}
\label{eq:Geg39}
\mathcal{C}^{(\alpha)}_{m}(z)\sim\frac{\alpha^{\frac12m}}{m!}H_m\left(z\sqrt{\alpha}\right),
\end{equation}
which corresponds to the limit in \eqref{eq:Geg27}. The next step is writing
\begin{equation}
\label{eq:Geg40}
h(t)=c_0+d_0t+t^2g_0(t),\quad c_0=h(0)=1,\quad d_0=h^\prime(0)=0,
\end{equation}
and substituting this in \eqref{eq:Geg38}. This gives
\begin{equation}
\label{eq:Geg41}
\begin{array}{@{}r@{\;}c@{\;}l@{}}
\mathcal{C}^{(\alpha)}_{m}(z)&=&\dsp{ \frac{\alpha^{\frac12m}}{m!}\left(c_0H_m\left(z\sqrt{\alpha}\right)+\frac{m}{\sqrt{\alpha}}d_0
H_{m-1}\left(z\sqrt{\alpha}\right)\right)\ +} \\[8pt]
&&\dsp{\frac{1}{2\pi i}\int_\calC e^{-\alpha\left(t^2-2zt\right)}g_0(t)\,\frac{dt}{t^{m-1}}.}
\end{array}
\end{equation}
Integrating by parts, writing $\dsp{e^{-\alpha t^2}\,dt=-\frac{1}{2\alpha t}\,d e^{-\alpha t^2}}$, we find
\begin{equation}
\label{eq:Geg42}
\begin{array}{@{}r@{\;}c@{\;}l@{}}
\mathcal{C}^{(\alpha)}_{m}(z)&=&\dsp{ \frac{\alpha^{\frac12m}}{m!}\left(c_0H_m\left(z\sqrt{\alpha}\right)+\frac{m}{\sqrt{\alpha}}d_0
H_{m-1}\left(z\sqrt{\alpha}\right)\right)\ +} \\[8pt]
&&\dsp{\frac{1}{2\pi i\alpha}\int_\calC e^{-\alpha\left(t^2-2zt\right)}h_1(t)\,\frac{dt}{t^{m+1}},}
\end{array}
\end{equation}
where
\begin{equation}
\label{eq:Geg43}
h_1(t)=\tfrac12 e^{-2z\alpha t}t^{m+1}\frac{d}{dt}\frac{e^{2z\alpha t}g_0(t)}{t^{m}}=\tfrac12\bigl(tg_0^\prime(t)+(2z\alpha t-m)g_0(t)\bigr).
\end{equation}
Repeating this procedure, we find an expansion of the form given in \eqref{eq:Geg33}.

\begin{example}\label{ex:E4}
For $x=0$ and even $m=2n$ we find from 
\begin{equation}
\label{eq:Geg44}
H_{2n}(0)=(-1)^n\frac{(2n)!}{n!},\quad \mathcal{C}^{(\alpha)}_{2n}(0)=(-1)^n\frac{\Gamma(\alpha+n)}{n!\,\Gamma(\alpha)},
\end{equation}
and from \eqref{eq:Geg33} we conclude that the first series should be an expansion of $\dsp{\frac{\Gamma(\alpha+n)}{\alpha^n\,\Gamma(\alpha)} }$. This function has the expansion
\begin{equation}
\label{eq:Geg45}
\frac{\Gamma(\alpha+n)}{\alpha^n\,\Gamma(\alpha)}\sim 1+\frac{n(n-1)}{2\alpha}+\ldots,
\end{equation}
which confirms the first values given in \eqref{eq:Geg36}.
\eoexample
\end{example}

Next we explain how the coefficients $D_k$ of expansion \eqref{eq:Geg30} can be obtained. The Gegenbauer polynomial in \eqref{eq:Geg25} has argument $x$ and we need the polynomial in terms of $y$. For $c=d=1$ the relation between $x$ and $y$ is given by (see also \eqref{eq:Geg09})
\begin{equation}
\label{eq:Geg46}
-\log\left(1-x^2\right)=\tfrac12y^2, \quad x=a_1y\sqrt{\frac{1-e^{-\frac12y^2}}{\frac12y^2}},
\end{equation}
where the square root is positive and  $a_1=\frac12\sqrt{2}$. We write 
\begin{equation}
\label{eq:Geg47}
x=z+h,\quad z=\tfrac12y\sqrt{2}, \quad h=a_3y^3+a_5y^5+\ldots,
\end{equation}
and the coefficients easily follow from \eqref{eq:Geg46}. We expand
\begin{equation}
\label{eq:Geg48}
\mathcal{C}^{(\alpha)}_{m}(x)=\sum_{k=0}^\infty \frac{h^k}{k!}\frac{d^k}{dx^k}\left.\mathcal{C}^{(\alpha)}_{m}(x)\right\vert_{x=z},
\end{equation}
and we obtain expansions of the derivatives from \eqref{eq:Geg33}. The derivatives of the Hermite polynomials can be written in terms of the polynomials used in \eqref{eq:Geg33}. The result of straightforward manipulations is a representation of the form
\begin{equation}
\label{eq:Geg49}
\mathcal{C}^{(\alpha)}_{m}(x)\sim\frac{\alpha^{\frac12m}}{m!}\left(H_m\left(y\sqrt{\alpha/2}\right)P+\frac{m}{\sqrt{\alpha}}
H_{m-1}\left(y\sqrt{\alpha/2}\right)Q\right),
\end{equation}
in which $P$ and $Q$ can be expanded in powers of $y$, with coefficients that are finite combinations of the coefficients $c_k$ and $d_k$ and their derivatives. These expansions should be multiplied by that of $f(y)$ given in \eqref{eq:Geg25}. Next, this new compound expansion should be squared and finally we need to evaluate integrals of the form
\begin{equation}
\label{eq:Geg50}
\int_{-\infty}^\infty e^{-t^2}t^j H_m(t)H_{m-k}(t)\,dt,
\end{equation}
with $k=0, 1$ and for even $j+k$. 
For example,
\begin{equation}
\label{eq:Geg51}
\begin{array}{@{}r@{\;}c@{\;}l@{}}
\dsp{\int_{-\infty}^\infty e^{-t^2}tH_m(t)H_{m-1}(t)\,dt}&=&2^m(m+1)!\,\sqrt{\pi},\\ [8pt]
\dsp{\int_{-\infty}^\infty e^{-t^2}t^2\,H_m(t)H_{m}(t)\,dt}&=&2^{m-1}(2m+1)m!\,\sqrt{\pi}.
\end{array}
\end{equation}

Because of the lengthy calculations, which are all quite straightforward with symbolic calculations,  we skip the details.

\section{Extended asymptotic expansions of Laguerre integrals}
\label{sec:more}
In this section we obtain the asymptotic expansion of the R\'{e}nyi and Shannon-like integral functionals of Laguerre polynomials given by Eqs. (\ref{eq:intro01}) and (\ref{eq:intro02}), respectively, for large positive values of the parameters $\alpha$ and $\mu$, while the other parameter $\lambda$ is fixed. We consider the R\'{e}nyi-like integral in \eqref{eq:intro01} for the case $\mu=\bigO(\alpha)$ in the special form $\mu=\sigma+\alpha$, with $\sigma$ a fixed real number.
%We need to distinguish between $\lambda=1$ and $\lambda\ne 1$.
%\subsection{The case $\lambda=1$}\label{sec:moresub1}

We write 
\begin{equation}
\label{eq:more01}
I_{5} (m,\alpha)= \int\limits_{0}^{\infty}x^{\alpha+\sigma-1}e^{-\lambda x} \left|\mathcal{L}_{m}^{(\alpha)}(x)\right|^{\kappa}\,dx,
\end{equation}
or
 \begin{equation}
\label{eq:more02}
I_{5} (m,\alpha)=\alpha^{\alpha+\sigma} e^{-\alpha}
\int\limits_{0}^{\infty}x^{\sigma-1}e^{-\alpha\phi(x)} \left|\mathcal{L}_{m}^{(\alpha)}(\alpha x)\right|^{\kappa}\,dx,
\end{equation}
where
 \begin{equation}
\label{eq:more03}
\phi(x)=\lambda x-\log x -1, \quad \phi^\prime(x)=\frac{\lambda x-1}{x}.
\end{equation}
There is a saddle point at $x_0=1/\lambda$ and we use the transformation
 \begin{equation}
\label{eq:more04}
\phi(x)-\phi(x_0)=\tfrac12y^2,\quad \sign(x-x_0)=\sign(y),\quad \phi(x_0)=\log\lambda.
\end{equation}
For small $y$ we have the expansion   \begin{equation}
\label{eq:more05}
x=x_0+\sum_{k=1}^\infty x_k y^k,
\end{equation}
with first coefficients
\begin{equation}
\label{eq:more06}
x_1=\frac{1}{\lambda},\quad x_2=\frac{1}{3\lambda},\quad x_3=\frac{1}{36\lambda},\quad x_4=-\frac{1}{270\lambda},\quad x_5=\frac{1}{4320\lambda}.
\end{equation}

The transformation in \eqref{eq:more04}  gives, because $\dsp{\frac{dx}{dy}=\frac{xy}{\lambda x-1}}$,
\begin{equation}
\label{eq:more07}
I_{5} (m,\alpha)=\alpha^{\alpha+\sigma} e^{-\alpha}\lambda^{-\alpha}
\int\limits_{-\infty}^{\infty}e^{-\frac12\alpha y^2} f(y)\,dy,
\end{equation}
where
\begin{equation}
\label{eq:more08}
f(y)=\frac{x^{\sigma}\,y}{\lambda x-1}\left|\mathcal{L}_{m}^{(\alpha)}(\alpha x)\right|^{\kappa}.
\end{equation}
and  the relation between $x$ and $y$ is defined in \eqref{eq:more04}.
We need to distinguish between $\lambda=1$ and $\lambda\ne 1$.

\subsection{The case $\lambda\ne 1$}\label{sec:moresub1}
We use the expansion given in \eqref{eq:Lag07} and write
\begin{equation}
\label{eq:more09}
f(y)=\frac{x^{\sigma}\,y}{\lambda x-1}
\frac{\alpha^{\kappa m}}{(m!)^\kappa} \vert1-x\vert^{\kappa m}\sum_{j=0}^\infty \frac{A_{j}}{\alpha^j}(1-x)^{-j}.
\end{equation}
Using the expansion given in \eqref{eq:more05} in the form
\begin{equation}
\label{eq:more10}
1-x=(1-x_0)\left(1-\frac{x_1y}{1-x_0}-\frac{x_2y^2}{1-x_0}-\ldots\right),
\end{equation}
we expand
\begin{equation}
\label{eq:more11}
f(y)\sim x_0^\sigma\left\vert 1-x_0\right\vert^{\kappa m}\frac{\alpha^{\kappa m}}{(m!)^\kappa}\sum_{j=0}^\infty C_j(\alpha)y^j,
\end{equation}
and obtain for the integral $I_{5} (m,\alpha)$ given in  \eqref{eq:more07} the expansion
\begin{equation}
\label{eq:more12}
I_{5} (m,\alpha)\sim\alpha^{\alpha+\sigma} e^{-\alpha}\lambda^{-\alpha}x_0^\sigma\left\vert 1-x_0\right\vert^{\kappa m}
\sqrt{\frac{2\pi}{\alpha}}\frac{\alpha^{\kappa m}}{(m!)^\kappa}\sum_{j=0}^\infty C_{2j}(\alpha)\frac{2^j\left(\frac12\right)_j}{\alpha^j}.
\end{equation}
We can expand the coefficients  $C_{2j}(\alpha)$ for large $\alpha$ and, as explained for Proposition~\ref{prop:P1},  we can rearrange the series to obtain an expansion in negative powers of $\alpha$. This gives the asymptotics
\begin{equation}
\label{eq:more13}
I_{5} (m,\alpha)\sim\alpha^{\alpha+\sigma} e^{-\alpha}\lambda^{-\alpha-\sigma-\kappa m}\left\vert \lambda-1\right\vert^{\kappa m}
\sqrt{\frac{2\pi}{\alpha}}\frac{\alpha^{\kappa m}}{(m!)^\kappa}\sum_{j=0}^\infty \frac{D_{j}}{\alpha^j},
\end{equation}
with first coefficients $D_0=1$ and 
\begin{equation}
\label{eq:more14}
\begin{array}{@{}r@{\;}c@{\;}l@{}}
D_1&=&\dsp{\frac{1}{12(\lambda-1)^2}\Bigl(1-12 \kappa m \sigma \lambda+6 \sigma^2 \lambda^2-12 \sigma^2 \lambda-6 \sigma \lambda^2+12 \sigma \lambda\ +}
\\ [8pt]
&&\quad\quad
6 \kappa^2 m^2+12 \kappa m \sigma-12 \kappa m^2 \lambda-12 \kappa m \lambda+6 \kappa m \lambda^2+\\ [8pt]
&&\quad\quad
6 \kappa m^2 \lambda^2+\lambda^2+6 \sigma^2-2 \lambda-6 \sigma+6 \kappa m^2\Bigr),
\end{array}
\end{equation}
for the R\'{e}nyi-like integral functional (\ref{eq:more01}) when $\alpha \to\infty$ and the rest of parameters ($\sigma, \lambda \ne 1, \kappa, m$) are fixed.
\begin{example}\label{ex:E5}
We have the special case for $\kappa=2$ and $\sigma=1$  (see \cite[Page~477]{Prudnikov})
\begin{equation}
\label{eq:more15}
\begin{array}{@{}r@{\;}c@{\;}l@{}}
I_{5} (m,\alpha)&=&\dsp{ \int\limits_{0}^{\infty}x^{\alpha}e^{-\lambda x} \mathcal{L}_{m}^{(\alpha)}(x)^{2}\,dx=
\frac{(\alpha+1)_m(\alpha+1)_m\Gamma(\alpha+1)(\lambda-1)^{2m}}{m!\,m!\,\lambda^{2m+\alpha+1}}\ \times}\\ [8pt]
&&\quad\quad\dsp{\ {}_2F_1\left(-m,-m;\alpha+1;\frac{1}{(\lambda-1)^2}\right).}
\end{array}
\end{equation}
We can expand  the Pochhammer symbols as in \eqref{eq:Lag19}, the gamma function, and the $\ {}_2F_1$-function term by term in negative powers of $\alpha$. We obtain
\begin{equation}
\label{eq:more16}
\begin{array}{@{}r@{\;}c@{\;}l@{}}
&&I_{5} (m,\alpha)\sim\dsp{\frac{\alpha^{2m+1+\alpha}e^{-\alpha}(\lambda-1)^{2m}}{m!\,m!\,\lambda^{2m+\alpha+1}}\sqrt{\frac{2\pi}{\alpha}}\ } \times\\ [8pt]
&&\quad\dsp{\left(1+\frac{m(m+1)}{2\alpha}+\ldots\right)^2\left(1+\frac{1}{12\alpha}+\ldots\right)\left(1+\frac{m^2}{\alpha(\lambda-1)^2}+\ldots\right).}
\end{array}
\end{equation}

We obtain the expansion as  in \eqref{eq:more13}, with the same front factor and coefficients $D_0=1$  and
\begin{equation}
\label{eq:more17}
D_{1}=  \frac{24m^2+\lambda^2-2\lambda+1+12m^2\lambda^2-24m^2\lambda+12m\lambda^2-24m\lambda+12m}{12(\lambda-1)^2},
\end{equation}
which confirms $D_1$ given in \eqref{eq:more14} when $\kappa=2$ and $\lambda=1$.
\eoexample
\end{example}

\begin{remark}\label{rem:rem05}
The Shannon-like integral 
\begin{equation}
\label{eq:more18}
I_{5} ^{*}(m,\alpha)= \int\limits_{0}^{\infty}x^{\alpha+\sigma-1}e^{-\lambda x} \bigl(\mathcal{L}_{m}^{(\alpha)}(x)\bigr)^{2}\,\log\bigl(\mathcal{L}_{m}^{(\alpha)}(x)\bigr)^{2}\,dx
\end{equation}
follows from differentiating \eqref{eq:more13} with respect to $\kappa$ and putting $\kappa=2$ afterwards.
The result is
\begin{equation}
\label{eq:more19}
\begin{array}{@{}r@{\;}c@{\;}l@{}}
I_{5} ^{*}(m,\alpha)&\sim&\dsp{\alpha^{\alpha+\sigma} e^{-\alpha}\lambda^{-\alpha-\sigma-2 m}( \lambda-1)^{2 m}
\sqrt{\frac{2\pi}{\alpha}}\frac{\alpha^{2 m}}{(m!)^2}\ \times}\\ [8pt]
&&
\dsp{\left(\log\frac{\alpha^{2m}(\lambda-1)^{2m}}{\lambda^{2m}(m!)^{2m}}\sum_{j=0}^\infty \frac{D_{j}}{\alpha^j}+2\sum_{j=0}^\infty\frac{D_{j}^\prime}{\alpha^j}\right),}
\end{array}
\end{equation}
where the derivatives of $D_j$ are with respect to $\kappa$.
\eoremark
\end{remark}

\subsection{The case $\lambda = 1$}\label{sec:moresub2}
The limit given in \eqref{eq:limit01} cannot be used in this case, because it does not give enough information in the immediate neighborhood of $x=1$, that is, $y=0$. The asymptotic relation in \eqref{eq:Lag02} that follows from the limit is not uniformly valid at $t=1$. A similar form of nonuniform behavior is considered in \S\ref{sec:disc} for the Gegenbauer polynomial.

In this case we use the limit (see \cite[Eqn.~18.7.26]{nist1})
\begin{equation}
\label{eq:more20}
\lim_{\alpha\to\infty}\left(\frac{2}{\alpha}\right)^{\frac12m}\mathcal{L}_{m}^{(\alpha)}\left(\sqrt{2\alpha}\,x+\alpha\right)=\frac{(-1)^m}{m!}H_m(x),
\end{equation}
and we write it as the asymptotic relation
\begin{equation}
\label{eq:more21}
\mathcal{L}_{m}^{(\alpha)}(\alpha x)\sim \left(\frac{\alpha}{2}\right)^{\frac12m}\frac{(-1)^m}{m!}H_m\left(\sqrt{\frac{\alpha}{2}}(x-1)\right).\end{equation}
We use this estimate in \eqref{eq:more06}, replace $x-1$ by $y$ and $x^{\sigma}\,y/(x-1)$ by 1, and obtain the first approximation
\begin{equation}
\label{eq:more22}
I_{5} (m,\alpha)\sim\alpha^{\alpha+\sigma} e^{-\alpha}\frac{1}{(m!)^\kappa}\left(\frac{\alpha}{2}\right)^{\frac12\kappa m}
\int\limits_{-\infty}^{\infty}e^{-\frac12\alpha y^2} \left|H_m\left(\sqrt{\frac{\alpha}{2}}\,y\right)\right|^{\kappa}\,dy,
\end{equation}
for the R\'{e}nyi-like integral functional (\ref{eq:more01}) when $\alpha \to\infty$ and the rest of parameters ($\sigma, \lambda = 1, \kappa, m$) are fixed.. In particular, for $\kappa=2$ we can evaluate this integral by using a standard result for the orthogonal Hermite polynomials; we finally have
\begin{equation}
\label{eq:more23}
I_{5} (m,\alpha)\sim\frac{\alpha^{\alpha+\sigma+m} e^{-\alpha}}{m!}
\sqrt{\frac{2\pi}{\alpha}}. 
\end{equation}
We can obtain more details of the asymptotic estimate by using an approach similar to that described for the Gegenbauer polynomials in \S\ref{sec:gegher}.

\begin{example}\label{ex:E6}
We take $\sigma=1$, and have, by the orthogonality relation of the Laguerre polynomials,
\begin{equation}
\label{eq:more24}
I_{5} (m,\alpha)= \int\limits_{0}^{\infty}x^{\alpha}e^{-x} \mathcal{L}_{m}^{(\alpha)}(x)^{2}\,dx=\frac{\Gamma(m+\alpha+1)}{m!}.
\end{equation}
The large-$\alpha$ asymptotic result in the right-hand side of \eqref{eq:more10} corresponds to that of \eqref{eq:more11}. It is easy to verify that this correspondence does not happen when we use the asymptotic relation in \eqref{eq:Lag02} instead of the one in \eqref{eq:more08}.
\eoexample
\end{example}

\begin{remark}\label{rem:rem06}
A result for the Shannon-like integral in \eqref{eq:more18} with $\lambda=1$
follows from differentiating \eqref{eq:more22} with respect to $\kappa$ and putting $\kappa=2$ afterwards. It seems not to be possible to give a large-$\alpha$ expansion of  the resulting integral.
\eoremark
\end{remark}

\subsection{Hermite-type expansion of the Laguerre polynomials}\label{sec:lagher}
We can find more asymptotic details of the approximation in \eqref{eq:more23} when we expand the Laguerre polynomials in an asymptotic representation in terms of the Hermite polynomials of the form
\begin{equation}
\label{eq:more25}
\begin{array}{@{}r@{\;}c@{\;}l@{}}
&&\dsp{\mathcal{L}_{m}^{(\alpha)}(\alpha x)\sim \left(\frac{\alpha}{2}\right)^{\frac12m}\frac{(-1)^m}{m!}\ \times} \\ [8pt]
&&
\dsp{\left(H_m\left(\sqrt{\frac{\alpha}{2}}(x-1)\right)\sum_{k=0}^\infty\frac{c_k}{\alpha^k}-m\sqrt{\frac{2}{\alpha}}
H_{m-1}\left(\sqrt{\frac{\alpha}{2}}(x-1)\right)\sum_{k=0}^\infty\frac{d_k}{\alpha^k}\right).}
\end{array}
\end{equation}
The first coefficients are
\begin{equation}
\label{eq:more26}
\begin{array}{@{}r@{\;}c@{\;}l@{}}
c_0&=&1,\quad d_0=1, \\ [8pt]
c_1&=&m\left(\alpha(x-1)-1\right),\\ [8pt]
d_1&=&\frac{1}{3}\left(3+7\alpha-3 m-9 \alpha x+3 \alpha^2(x-1)^2-4 \alpha m+6 \alpha x m\right).
\end{array}
\end{equation}

The expansion is valid for bounded values of the argument of the Hermite polynomials. Information on Hermite-type uniform expansions for large $\alpha$ and degree $m$ can be found in \cite{Temme:1990:AEL} and in \cite[\S32.4]{temme2}.   

The expansion in \eqref{eq:more25} can be derived by using the well-known integral representation
\begin{equation}
\label{eq:more27}
\mathcal{L}_{m}^{(\alpha)}(x)=\frac{1}{2\pi i}\int_{\calC}(1-t)^{-\alpha-1}e^{-xt/(1-t}\,\frac{dt}{t^{n+1}}
\end{equation}
where $\calC$ is a circle with radius 1 around the origin. We write this in the form
\begin{equation}
\label{eq:more28}
\mathcal{L}_{m}^{(\alpha)}(\alpha x)=\frac{1}{2\pi i}\int_{\calC}h(t)e^{\alpha(1-x)t-\frac12\alpha t^2}\,\frac{dt}{t^{n+1}},
\end{equation}
where 
\begin{equation}
\label{eq:more29}
h(t)=\frac{1}{ (1-t)}e^{\alpha\left(-\log(1-t)-xt/(1-t)-(1-x)t+\frac12 t^2\right)}.
\end{equation}
Integrating by parts, starting with $\dsp{e^{\frac12\alpha t^2}\,dt=\frac{1}{\alpha t}\,de^{\frac12\alpha t^2}}$, and using the procedure described in \S\ref{sec:gegher}, we can obtain the expansion given in \eqref{eq:more25}.

Next, we need to expand this expansion in terms of $y$ (see \eqref{eq:more05} with $x_0=1$ and $\lambda=1$), and we can obtain more details of the asymptotic relation in  \eqref{eq:more23} when we use the method used for the Gegenbauer polynomials. Again, see  
 \S\ref{sec:gegher}. We skip further details.
 
 \section{Concluding remarks}
 \label{sec:concl}
 We have investigated in a detailed manner the asymptotics of the power and logarithmic integral functionals of Laguerre and Gegenbauer polynomials $I_{j} (m,\alpha), j= 1-4$ when the parameter $\alpha \to\infty$ and the rest of parameters, including the polynomial degree $m$, are fixed. These asymptotic functionals of power and logarithmic kind characterize the R\'{e}nyi and Shannon entropies, respectively, of numerous quantum systems with a large dimensionality $D$.
 
 Because of the many parameters in some of the integrals only a limited number of coefficients of the expansions have been given, and we have used special cases of these variables as examples for which analytic closed forms can be found in the literature. We have used these analytic forms to show how their simple asymptotic results correspond to the derived expansions, and confirm our (in some cases rather formal) approach. 
 
 We have derived new Hermite-type expansions of the Laguerre and Gegenbauer polynomials, which in fact are not as powerful as known uniform expansions, but the expansions are useful in the analysis of certain special cases of the functionals. 
 
 As always, certain problems remain to be studied. We have not been able to determine the large-$\alpha$ expansion of the Shannon-like integral in \eqref{eq:more18} with $\lambda=1$ within our approach. Moreover, the determination of the asymptotics of the Laguerre and Gegenbauer polynomials for large values of the degree of the polynomials is most important from both fundamental and applied standpoints. Indeed, it would allow for the analytical calculation of the physical entropies of the highly-excited (i.e., Rydberg) states of numerous hydrogenic and harmonic systems. Let us just point out that the underlying asymptotic analysis for large degree is essentially more difficult than the large-parameter case studied in the present paper and it requires other asymptotical tools. Nevertheless, some remarkable results have been obtained \cite{aptekarev,apteka2, buyarov,tor2016a,tor2016b} (see also the reviews \cite{dehesa2001,apteka1}).

\section*{Acknowledgments}
This work has been partially supported by the Projects FQM-7276 and FQM-207 of the Junta de Andaluc\'ia and the MINECO-FEDER grants
 FIS2014- 54497P and FIS2014-59311-P. I. V. Toranzo acknowledges the support of ME under the program FPU. 
 N. M. Temme acknowledges financial support from 
{\emph{Ministerio de Ciencia e Innovaci\'on}}, project MTM2012-11686,  and thanks CWI, Amsterdam, for scientific support.

%\section*{References}

\end{document}